\documentclass[%
 reprint, 
 superscriptaddress,
 amsmath,amssymb,
 aps, physrev,
]{revtex4-2}

\usepackage{graphicx}
\usepackage{xcolor}
\usepackage{dcolumn}
\usepackage{bm}
\usepackage[hidelinks]{hyperref}
\usepackage[percent]{overpic}

\begin{document}

\preprint{APS/123-QED}

\title{\textbf{Polarized positron generation via laser-engineered quasistatic magnetic fields in a spatially extended herringbone target
    }}

\author{Xian-Zhang Wu}
\affiliation{Department of Nuclear Science and Technology, Xi'an Jiaotong University, Xi'an, China}

\author{Hui-Jing Li}
\affiliation{Department of Nuclear Science and Technology, Xi'an Jiaotong University, Xi'an, China}

\author{Huai-Hang Song}
\affiliation{School of Physics, Zhengzhou University, Henan, China}

\author{Wei-Min Wang}
\email{weiminwang1@sjtu.edu.cn}
\affiliation{School of Physics and Astronomy, Shanghai Jiao Tong University, Shanghai, China}

\author{Yan-Fei Li}
\email{liyanfei@xjtu.edu.cn}
\affiliation{Department of Nuclear Science and Technology, Xi'an Jiaotong University, Xi'an, China}

\date{\today}

\begin{abstract}

Polarized positron sources are of interest for precision studies of spin-dependent interactions and future high-energy colliders, yet their laser-driven realization remains challenging because of demanding laser or beam requirements. Here, we propose a target-engineered scheme for generating polarized positrons using a 3.4-PW single-color laser pulse and an initially unpolarized 10-GeV electron beam. 
Three-dimensional spin-resolved quantum electrodynamics particle-in-cell simulations show that laser-driven return currents in a herringbone target generate spatially extended quasistatic magnetic fields exceeding $\simeq 10^6$ T, with opposite polarities and a duration exceeding 100 fs. 
These fields correlate positron spin polarization with deflection direction, enabling angular selection of two oppositely polarized positron populations with polarization magnitudes of 63\% and a combined yield of approximately 0.007 positrons per incident electron, compared with a total yield of 0.029. Our results demonstrate a route toward highly polarized positron generation under laser conditions accessible to existing petawatt-class facilities, without requiring frequency-converted ultraintense fields or prepolarized incident beams.

\end{abstract}

\maketitle

\newpage
\section{Introduction}

\begin{figure*}[t]
    \centering
    \begin{overpic}[width=0.8\textwidth]{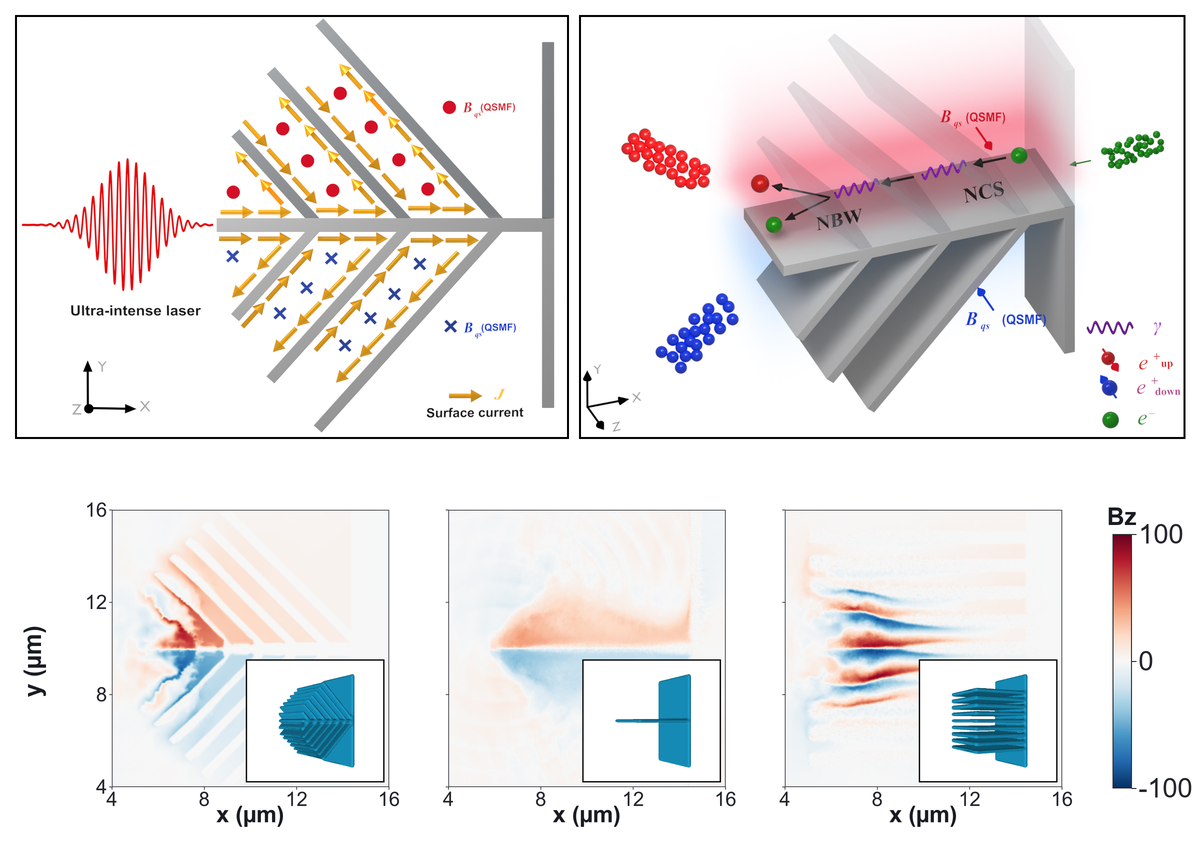}
        \put(1.5,65.0){\colorbox{white}{\textbf{\large (a)}}}
        \put(48.5,65.0){\colorbox{white}{\textbf{\large (b)}}}
        \put(10,25){\colorbox{white}{\textbf{ (c1)}}}
        \put(38,25.0){\colorbox{white}{\textbf{ (c2)}}}
        \put(65.5,25.0){\colorbox{white}{\textbf{(c3)}}}
    \end{overpic}
    \caption{
        Schematic of the polarized-positron generation scheme and comparison of target configurations. (a) Laser irradiation of the herringbone target generates surface return currents and quasi-static magnetic fields. (b) A delayed relativistic electron beam produces high-energy photons through nonlinear Compton scattering, followed by polarized positron production via the nonlinear Breit-Wheeler process. The opposite magnetic-field polarities lead to spatial separation of the positrons. (c1-c3) Quasistatic magnetic-field distributions in the \(x\)–\(y\) plane at \(z=10\,\mu\mathrm{m}\) for the herringbone, planar-foil, and multiple-foil targets, respectively, under identical laser conditions, $t=36T_0$.
        }
    \label{fig:scheme}
\end{figure*}

Polarized positron beams are valuable probes of spin-dependent
interactions in particle and nuclear physics and have potential
applications in precision measurements and future high-energy
colliders \cite{Moortgat2008,Mane2005,baier1971radiative}.
In particular, positron polarization can provide additional
sensitivity to electroweak interactions and improve the physics
reach of polarized electron-positron collisions. Conventional
polarized positron sources generally rely on the production of
polarized photons followed by pair conversion, while radiative
polarization in storage rings provides an alternative based on
spin-dependent synchrotron radiation
\cite{Sokolov1964,baier1971radiative,Derbenev_1973,Jackson1976,Ternov1995}.
Although these approaches are well established,
their implementation involves substantial accelerator
infrastructure and, in the case of radiative polarization,
relatively long polarization times. These considerations have
motivated the development of laser-driven approaches that can
generate relativistic positrons with useful polarization on
ultrashort timescales, offering a complementary route to
conventional accelerator-based methods.

Ultraintense lasers provide a platform for developing
such sources because they can accelerate electrons and
generate high-energy photons on femtosecond timescales,
bringing nonlinear quantum electrodynamics (QED) processes
into the interaction \cite{Piazza2012,Ritus1985,Nikishov1964,Nikishov1967}.
In particular, nonlinear Compton scattering (NCS) and nonlinear
Breit-Wheeler (NBW) pair production provide an efficient
sequence for converting the energy of relativistic electrons
and intense electromagnetic fields into high-energy photons
and electron-positron pairs
\cite{Breit1934,burke1997positron,bula1996observation,bamber1999studies}.
The spin dependence of these processes also provides a means of
generating and controlling polarized leptons. Theoretical
studies have shown that spin-dependent radiation emission
can polarize relativistic electrons in strong laser fields,
establishing a laser-driven analogue of radiative polarization
\cite{DelSorbo2017,DelSorbo2018,Seipt2018}. These developments
have stimulated theoretical proposals for producing polarized
positrons directly through strong-field QED processes.

A challenge in laser-driven polarized positron production arises from the oscillatory and approximately symmetric nature of the driving laser field.
Over successive optical cycles, the field reverses its direction,
so that spin-dependent asymmetries associated with the instantaneous
field can partially cancel when the interaction is integrated over
the pulse \cite{Liyf2019}. Producing a net positron polarization
therefore generally requires an additional source of asymmetry or
polarization. One approach is to break the temporal symmetry of
the driving field, for example by using two-color laser pulses,
which can generate polarized positrons through the spin asymmetry
of the NBW process \cite{PhysRevLett.123.174801}. Related schemes
exploit tailored laser polarization or other field configurations
to control the polarization transferred during photon emission
and pair production
\cite{Wan2020,Xue2022ArbitrarilyPolarized,Tang2022FullyPolarized}.
A different approach is to introduce the required spin asymmetry
through the incident particles. A longitudinally polarized electron
beam can transfer its helicity to the emitted photons and subsequently
to the positrons produced through NBW, allowing polarized positrons
to be generated even with a circularly polarized laser field
\cite{li2022helicity}. These studies establish that controlling
the symmetry and polarization of the electromagnetic environment,
or supplying a polarized initial state, is central to laser-driven
polarized positron production.

\begin{figure*}[htpb]
    \centering
    \includegraphics[width=1.0\textwidth]{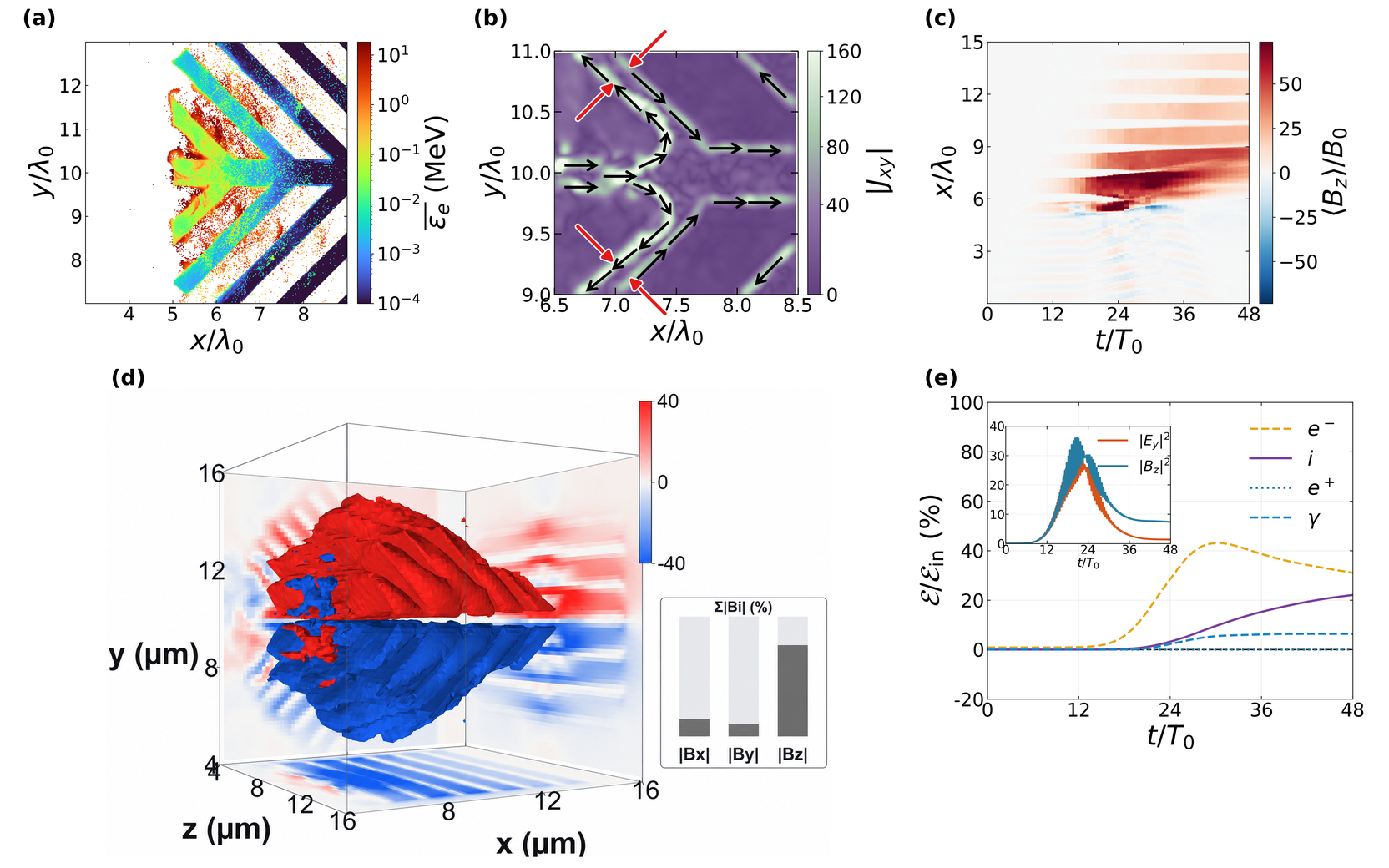}
    \caption{
        Generation and evolution of the quasi-static magnetic field. 
        (a) Spatial distribution of the mean electron kinetic energy, $\overline{\varepsilon_e}$, within the central slab, $9.5<z/\lambda_0<10.5$, at $t=12T_0$.
        (b) In-plane current density magnitude, $|\mathbf{J}_{xy}|=(J_x^2+J_y^2)^{1/2}$, in the $z=10\lambda_0$ plane at $t=24T_0$. Black and red arrows indicate the current direction and magnetic pinching force, respectively. 
        (c) Spatiotemporal evolution of the cycle-averaged magnetic field $\langle B_z\rangle$ at $y\simeq10.3\lambda_0$ and $z=10\lambda_0$.
        (d) Three-dimensional distribution of the cycle-averaged magnetic field $\langle B_z\rangle$ at $t=36T_0$. The bar chart shows the normalized contributions of $\sum|B_x|$, $\sum|B_y|$, and $\sum|B_z|$.
        (e) Temporal evolution of the particle and electromagnetic-field energies normalized to the cumulative laser input energy. The inset shows the normalized energies stored in the $E_y$ and $B_z$ components.
    }
    \label{fig:magnetic_field}
\end{figure*}

Plasma dynamics provide an alternative route to overcoming this limitation. Unlike the rapidly oscillating electromagnetic field of the incident laser, laser-driven plasma currents can generate magnetic fields that evolve on timescales much longer than an optical cycle. These self-generated quasi-static magnetic fields arise naturally from relativistic electron transport, return currents, plasma-density gradients, and other collective processes, and can reach strengths far exceeding those of conventional laboratory magnets \cite{Stamper_1991,Cai2007,Tatarakis2002,Santos2015}. Such fields can maintain a directional magnetic background within the interaction region, thereby providing a persistent spin-dependent environment for subsequent QED radiation and pair-production processes. 
Consequently, the use of plasma-generated fields for lepton polarization has therefore attracted increasing attention \cite{PhysRevLett.129.035001}. For example, a dense relativistic electron beam can drive a strong return current inside a target, generating asymmetric quasi-static magnetic fields and dense multi-GeV positron beams with average polarizations exceeding \(40\%\) \cite{PhysRevLett.132.235001}. Asymmetric quasi-static magnetic fields generated through tilted-foil laser--solid interactions can also produce polarized positrons, with an average polarization of approximately \(70\%\) at a laser intensity of \(10^{23}\,\mathrm{W/cm^2}\) \cite{PhysRevLett.131.175101}. In addition, target-normal sheath fields generated by laser-driven
plasma expansion have been proposed to produce positron beams with
polarizations of up to \(49\%\) using \(20~\mathrm{PW}\) laser
\cite{Liu2026PolarizedLepton}. 
Representative laser–plasma schemes basically require 100 PW-class driving lasers, while laser–electron beam collision schemes also face substantial experimental barriers. These requirements highlight the difficulty of achieving high positron polarization and appreciable yields under experimentally accessible conditions.


Here, we present an experimentally feasible scheme that operates within
the power capabilities of existing laser facilities, as illustrated in Fig.~\ref{fig:scheme}. An intense laser
pulse(3.4PW) irradiates the target, driving surface return currents that generate
a strong, spatially extended quasistatic magnetic field with opposite
polarities on the two sides of the target. The field exceeds $\simeq 10^6$
T and persists for more than 100 fs. A delayed relativistic unpolarized electron beam
is then injected into the magnetic-field region, where high-energy
photons are emitted through NCS and subsequently produce polarized
electron-positron pairs through the NBW.
Three-dimensional spin-resolved QED particle-in-cell simulations~\cite{wu2026spire} demonstrate the generation of two spatially separated positron populations with opposite spin polarizations. Angular selection yields a positron population with approximately 64\% polarization at 0.0037 positrons per incident electron. This polarization magnitude exceeds the 60\% level envisaged for a future International Linear Collider (ILC) upgrade \cite{Abramowicz2026LinearColliderVision}, although the spin orientation and beam-quality requirements for collider applications remain to be addressed.

\section{Results}

\subsection{Quasi-Static Magnetic-Field Generation and Characteristics}


Figures 1(c1)-(c3) compare the quasistatic magnetic fields generated by the herringbone, planar-foil, and multiple-foil targets under identical laser conditions. Compared with the other configurations, the herringbone target produces a spatially extended magnetic field with a well-organized polarity distribution. The underlying current dynamics are examined below.

Figure~\ref{fig:magnetic_field} presents multiscale diagnostics of the
laser--plasma interaction with the herringbone target.
During the early stage
of the interaction, the incident laser pulse has a normalized vector
potential of \(a_0=200\), corresponding to a relativistically corrected
critical density of
$
    n_{\mathrm{rel}}
    =
    \langle\gamma\rangle n_c
    \simeq
    \sqrt{1+\frac{a_0^2}{2}}\,n_c
    \simeq
    141\,n_c .
$
This value remains well below the plasma electron density,
\(n_e=900\,n_c\), where
\(n_c=1.1\times10^{21}~\mathrm{cm}^{-3}\) is the classical critical
density for a laser wavelength of \(\lambda_0=1~\mu\mathrm{m}\).
Consequently, the laser pulse cannot penetrate deeply into the overdense plasma and is predominantly reflected, while relativistic electrons are accelerated near the target surface.

Nevertheless, electrons at the target surface can be accelerated to
relativistic energies. Figure~\ref{fig:magnetic_field}(a) shows the
electron-density mean kinetic-energy distribution at
\(t=12T_0\). The surface electrons acquire substantially higher energies
than the cold electrons in the target interior and escape into the
surrounding vacuum and the gaps between the target branches. The
relativistic collisionless skin depth is estimated as
$
    \delta_{s,\mathrm{rel}}
    =
    \frac{c\sqrt{\langle\gamma\rangle}}{\omega_{pe}}
    =
    \frac{\lambda_0}{2\pi}
    \sqrt{\frac{\langle\gamma\rangle n_c}{n_e}}
    \simeq
    63~\mathrm{nm}.
$
The small relativistic skin depth implies that the return current, which is
established to compensate for the charge imbalance caused by the escape of
surface electrons, is confined primarily to a thin layer near the target
surface. As shown in Fig.~\ref{fig:magnetic_field}(b), at \(t=24T_0\), the
return current is concentrated along the surface of the structured target
and flows approximately parallel to the local target boundary. Within the
gaps between adjacent branches, these surface currents form a series of
approximately closed, loop-like current circuits with a well-organized
spatial topology. Consequently, they generate quasi-static magnetic fields
of nearly equal magnitude but opposite polarity in the regions above and
below the target. The peak return-current density reaches
$
    J_{\max}
    \simeq
    8.5\times10^{18}~\mathrm{A\,m^{-2}}.
$
Neglecting the displacement-current contribution, Ampère's law gives the
following order-of-magnitude estimate for the resulting quasi-static
magnetic field:
$
    B_{\mathrm{qs}}
    \sim
    \mu_0 K
    \sim
    \mu_0 J\delta_s ,
$
where \(K\sim J\delta_s\) denotes the effective surface current density.
This estimate yields \(B_{\mathrm{qs}}\sim 6.8\times10^{5}~\mathrm{T}\), in good
agreement with the magnetic-field strength obtained from the simulation.

The generated quasi-static magnetic field, in turn, acts back on the target
current and structure. As indicated by the red arrows in
Fig.~\ref{fig:magnetic_field}(b), the magnetic force density, or
Ampère force density,
$
    \boldsymbol{f}_{\mathrm A}
    =
    \boldsymbol{J}\times\boldsymbol{B}_{\mathrm{qs}},
$
is directed toward the central axis of each target branch.
The inward-directed magnetic force can compress the current layer, potentially increasing the local current density and contributing to the buildup of the quasistatic magnetic field. Similar current-driven pinching effects have been reported in laser-driven Z-pinch and wire-target configurations \cite{Kaymak2016,Beg2004WireReturnCurrent,
    Yang2024ReturnCurrentCompression,wang2018collimated}.

The inclined branches extend the return-current pathways in the transverse direction, increasing the characteristic thickness of the magnetic-field region from approximately \(0.1\,\mu\mathrm{m}\) in conventional planar-foil or wire configurations to the micrometer scale in the herringbone target.


The three-dimensional distribution of the quasi-static \(B_z\) field
generated by the herringbone target is shown in
Fig.~\ref{fig:magnetic_field}(d). The region enclosed by the isosurfaces
$
    \left|B_z\right|
    =
    10B_0
    =
    10\frac{m_e\omega_0}{e}
    \simeq
    1.07\times10^5~\mathrm{T}
$
extends approximately \(10~\mu\mathrm{m}\) in the transverse direction and
\(5~\mu\mathrm{m}\) in the longitudinal direction, demonstrating the large
spatial coverage of the generated magnetic field. In three-dimensional
geometry, the finite transverse size of the laser pulse inevitably drives a
non-negligible current component along the \(z\) direction, which in turn
generates magnetic-field components along the \(x\) and \(y\) directions.
Nevertheless, the statistical analysis shown in the lower-right inset of
Fig.~\ref{fig:magnetic_field}(d) indicates that the \(B_z\) component
contains the dominant fraction of the integrated field-magnitude contributions.
Therefore, the generated quasi-static magnetic field remains predominantly
oriented along the \(z\) direction.

Another important feature of the generated quasi-static magnetic field is
its long-term stability.  
As shown in Fig. 2(c), the field maintains a strength of approximately \(B_z\simeq50B_0\) for more than 100 fs, substantially longer than the duration of the driving laser pulse. The gaps between adjacent branches allow laser interaction with multiple internal surfaces, which contributes to sustained return currents and the extended magnetic-field lifetime.

Figure~\ref{fig:magnetic_field}(e) shows the temporal evolution of the energy
partition among the different particle species and electromagnetic fields.
At the beginning of the interaction, the plasma electrons absorb a large
fraction of the incident laser energy, resulting in a rapid increase in
their total energy. The electron energy reaches its maximum at
\(t=29T_0\), accounting for approximately \(29\%\) of the total input
energy. 
The electron energy subsequently decreases as energy is transferred to radiation, ions, and electromagnetic fields during the continued plasma evolution.

Although the plasma electrons absorb a substantial fraction of the laser
energy, the target configuration does not provide favorable conditions for
their sustained acceleration. Their maximum energy is limited to
approximately \(250~\mathrm{MeV}\). The emitted \(\gamma\)-ray photons carry
approximately \(10\%\) of the total input energy; however, under the present
interaction conditions, these photons do not produce an appreciable
positron yield. An additional high-energy electron beam is therefore
injected to generate sufficiently energetic \(\gamma\)-ray photons and
subsequently drive electron--positron pair production.

By \(t=34T_0\), the incident electromagnetic wave has been almost completely
reflected or has left the interaction region. The remaining electromagnetic
energy is then dominated by the quasi-static magnetic field and accounts for
approximately \(7\%\) of the total input energy (see Fig.~\ref{fig:magnetic_field}(e)). This fraction remains
nearly unchanged during the subsequent evolution, further demonstrating the
long-term stability of the generated quasi-static magnetic field.

\begin{figure*}[t]
    \centering
    \includegraphics[width=0.99\textwidth]{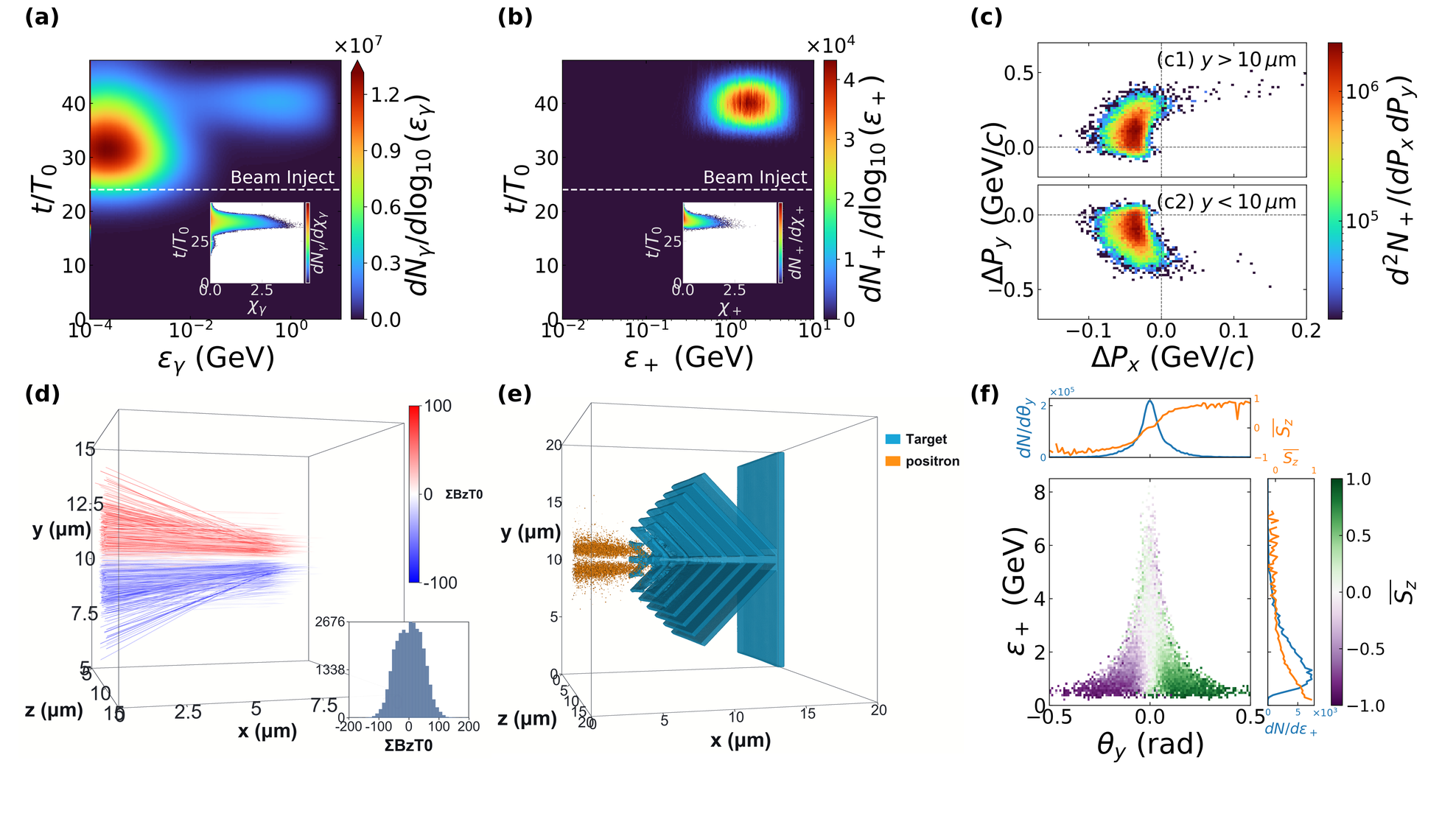}
    \caption{
        Temporal evolution and phase-space characteristics of the generated photons and positrons.
        (a,b) Time-resolved energy spectra of photons and positrons, respectively. The insets show the corresponding distributions of the quantum nonlinearity parameters $\chi_\gamma$ and $\chi_+$. The white dashed lines indicate the injection time of the electron beam.
        (c) Positron momentum-change distributions in the $(\Delta P_x,\Delta P_y)$ plane for the regions $y>10~\mu\mathrm{m}$ and $y<10~\mu\mathrm{m}$.
        (d) Representative three-dimensional positron trajectories colored by the accumulated magnetic field $\int B_z\,dt$ experienced by each particle. The inset shows the corresponding distribution of the accumulated field.
        (e) Spatial distribution of the positrons relative to the structured target.
        (f) Positron energy--angle distribution colored by the mean spin polarization $\langle S_z\rangle$. The upper and side panels show the angular and energy spectra, respectively, together with their corresponding mean spin polarizations.
    }
    \label{fig:photon_positron_generation}
\end{figure*}

\begin{figure*}
    \centering
    \includegraphics[width=0.7\textwidth]{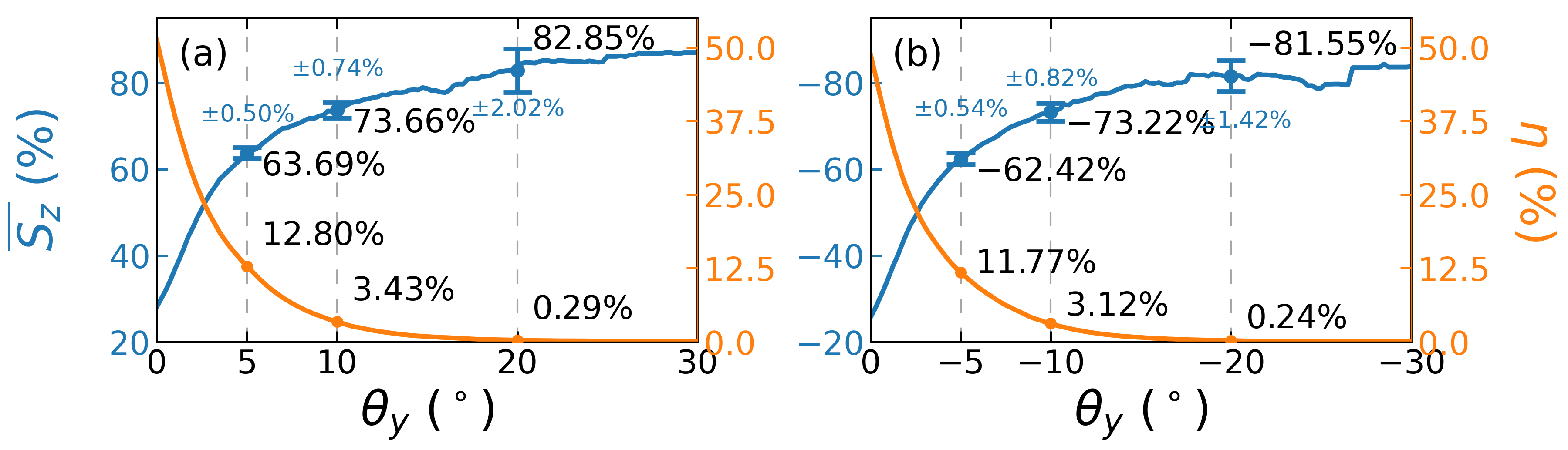}
    \caption{
Positron polarization $\overline{S_z}$ and relative yield
$\eta=N_+^{\mathrm{selected}}/N_+^{\mathrm{total}}$ as functions of the
angular-selection threshold $\theta_y$ for (a) $p_y>0$ and (b) $p_y<0$.
The dashed lines mark $|\theta_y|=5^\circ$, $10^\circ$, and $20^\circ$.
The error bars denote the weighted standard error.
}
    \label{fig:positron_polarization_yield}
\end{figure*}

\subsection{Radiation Emission and Positron Generation}


For target electrons with energies up to approximately 250 MeV, the quantum parameter is only \(\chi_e\simeq0.12\) at \(B_z=100B_0\), resulting in negligible pair production. We therefore inject a counterpropagating 10-GeV electron beam with \(\chi_e\simeq4.75\). The beam electrons emit high-energy photons through nonlinear Compton scattering, which subsequently produce electron-positron pairs via the nonlinear Breit-Wheeler process. Figures  ~\ref{fig:photon_positron_generation}(a) and  ~\ref{fig:photon_positron_generation}(b) show the temporal evolution of the photon and positron energy spectra, respectively.

The opposite magnetic-field polarities deflect positrons in opposite transverse directions, as shown by the momentum changes in Fig.  ~\ref{fig:photon_positron_generation}(c) and representative trajectories in Fig.  ~\ref{fig:photon_positron_generation}(d). Combined with the correlation between positron spin and the local magnetic field discussed in Sec. II.C, this produces spatially separated positron populations with opposite polarizations.


The simulation produces approximately \(1.799\times10^7\) physical positrons,
corresponding to a positron yield of 0.0288 per incident electron for the 0.1-nC
electron bunch. The two deflection branches have opposite spin-polarization
orientations, and their contributions largely cancel when integrated over the
entire positron population. Consequently, the net polarization of all generated
positrons is only +1.83\%, highlighting the importance of spatial or phase-space
selection for obtaining a positron beam with substantial net polarization.
Figure~\ref{fig:positron_polarization_yield} shows the mean spin polarization
and relative yield of positrons under different selection conditions. For an angular-selection threshold of \(|\theta_y|>5^\circ\), the \(p_y>0\) and \(p_y<0\) branches exhibit mean spin polarizations of \(+63.69\%\) and \(-62.42\%\), respectively. The selected populations account for 12.80\% and 11.77\% of the total positron yield, corresponding to approximately 0.0037 and 0.0034 positrons per incident electron. Thus, the two oppositely polarized branches provide a combined yield of approximately 0.0071 positrons per incident electron. Increasing the angular-selection threshold further enhances the polarization magnitude, but reduces the number of selected positrons. For
example, selecting positrons with \(|\theta_y|>10^\circ\) increases
the mean polarization to approximately 73\%, while the corresponding yield
decreases to only 3.28\%.

\par\vspace{\baselineskip}
\par\vspace{\baselineskip}

\begin{figure*}[htbp]
    \centering
    \includegraphics[width=0.99\textwidth]{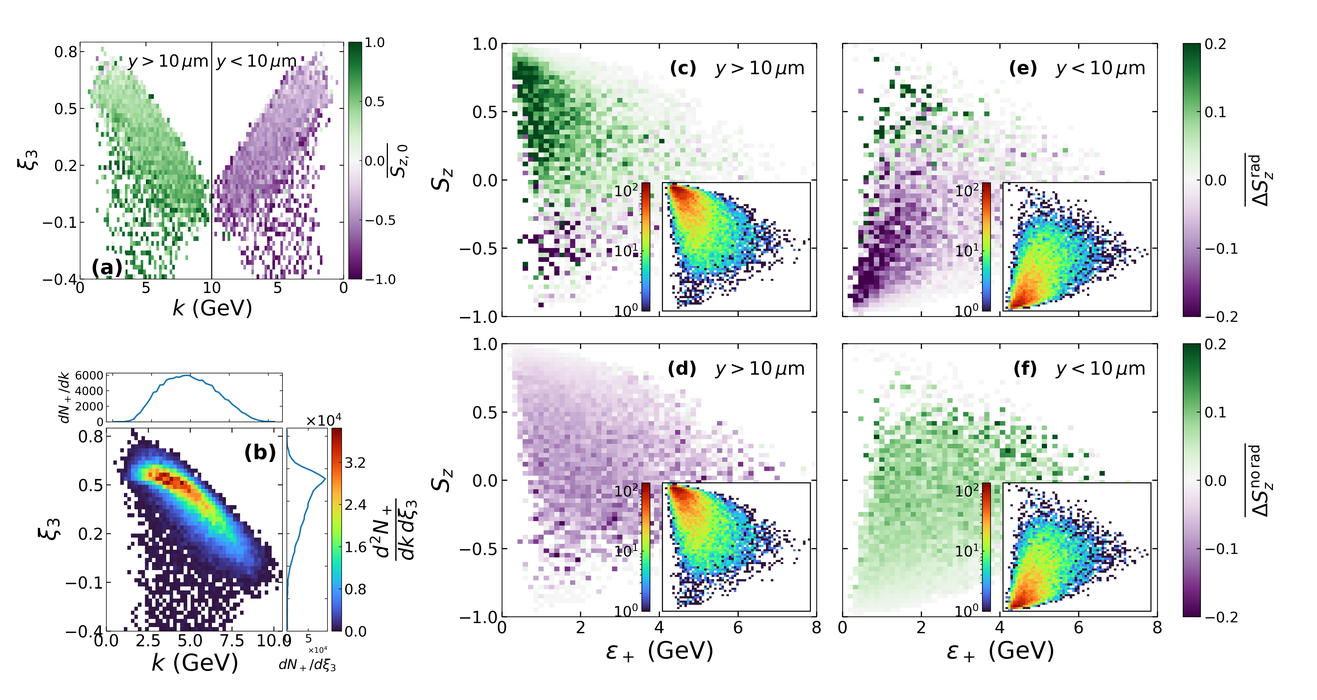}
    \caption{ Initial positron polarization and subsequent spin evolution. (a,b) Distributions of the initial spin polarization \(\overline{S_{z,0}}\) (a) and positron number density (b) in the parent-photon energy \(k\) and Stokes parameter \(\xi_3\) plane. The two halves of (a) correspond to \(y>10\,\mu\mathrm{m}\) and \(y<10\,\mu\mathrm{m}\), respectively. (c-f) Mean spin changes, in the (($\varepsilon_+,S_z$)) plane, due to radiative (c,e) and nonradiative (d,f) effects. Panels (c,d) and (e,f) correspond to $y>10 \mu\mathrm{m}$ and $y<10 \mu\mathrm{m}$, respectively. Insets show the corresponding positron-number distributions.
    }
    \label{fig:positron_spin_dynamics}
\end{figure*}

\subsection{Positron Spin Dynamics}

Four main processes can affect the spin polarization of the positrons:
(i) the initial spin polarization determined by the polarization of the
parent photons;
(ii) spin transitions associated with subsequent photon-emission events;
(iii) nonradiative spin evolution induced by the self-energy correction;
and (iv) spin precession.

According to the spin- and polarization-resolved probability of nonlinear
Breit--Wheeler pair production~\cite{Seipt2020,Chen2022,Song2021SpinPolarization},
the mean spin-polarization vector of the generated positrons is given by

\begin{widetext}
    \begin{equation}
        \mathbf{S}_{+}
        =
        \frac{
            -\xi_{1}\dfrac{\omega_{\gamma}}{\varepsilon_{-}}K_{1/3}(\rho)\hat{\mathbf e}_{1}
            -
            \left(
            \dfrac{\omega_{\gamma}}{\varepsilon_{+}}
            -
            \xi_{3}\dfrac{\omega_{\gamma}}{\varepsilon_{-}}
            \right)
            K_{1/3}(\rho)\hat{\mathbf e}_{2}
            +
            \xi_{2}
            \left[
                \dfrac{\omega_{\gamma}}{\varepsilon_{+}}\operatorname{Int}K_{1/3}(\rho)
                +
                \dfrac{\varepsilon_{+}^{2}-\varepsilon_{-}^{2}}
                {\varepsilon_{+}\varepsilon_{-}}
                K_{2/3}(\rho)
                \right]\hat{\mathbf e}_{v}
        }{
            \left(
            \dfrac{\omega_{\gamma}^{2}}
            {\varepsilon_{+}\varepsilon_{-}}
            -2
            \right)
            K_{2/3}(\rho)
            +
            \operatorname{Int}K_{1/3}(\rho)
            -
            \xi_{3}K_{2/3}(\rho).
        }
        \label{equation:nbw_spin}
    \end{equation}
\end{widetext}
Here, \(\hat{\mathbf e}_1\) is the unit vector along the positron acceleration,
\(\mathbf E+\hat{\mathbf e}_v\times\mathbf B-\hat{\mathbf e}_v(\hat{\mathbf e}_v
\cdot\mathbf E)\), where \(\hat{\mathbf e}_v=\mathbf v/|\mathbf v|\), and
\(\hat{\mathbf e}_2=\hat{\mathbf e}_v\times\hat{\mathbf e}_1\). The positron
and electron energies satisfy \(\varepsilon_++\varepsilon_-=\omega_\gamma\),
where \(\omega_\gamma\) is the incident photon energy. The parameters
\(\xi_{1,2,3}\) are the photon Stokes parameters, \(K_\nu\) denotes the
modified Bessel function of the second kind, and \(\operatorname{Int}K_{1/3}
(\rho)=\int_\rho^\infty K_{1/3}(x)\,\mathrm dx\),
with \(\rho=2\omega_\gamma^2/(3\chi_\gamma\varepsilon_+\varepsilon_-)\).
For photons emitted by initially unpolarized electrons, the average
values of \(\xi_1\) and \(\xi_2\) are approximately zero
\cite{Song2025PhotonPolarization,Song2021SpinPolarization}.
Since the denominator of Eq.~\eqref{equation:nbw_spin} is positive,
the sign of the positron spin component along \(\hat{\mathbf e}_2\) is
determined by \(1-\xi_3\varepsilon_+/\varepsilon_-\).



For $\chi_\gamma\simeq1$, the NBW process preferentially produces
electron--positron pairs with comparable energies~\cite{Seipt2020}, such that
$\varepsilon_+/\varepsilon_-\sim1$. Consequently, for most parent
photons with $\xi_3<1$, the coefficient
$1-\xi_3\varepsilon_+/\varepsilon_-$ is positive, and the initial positron spin
is preferentially aligned antiparallel to
$\hat{\mathbf e}_2$. In the magnetically dominated configuration, the
basis vectors employed in the code satisfy
$\hat{\mathbf e}_1\parallel\hat{\mathbf e}_v\times\mathbf B$ and
$
    \hat{\mathbf e}_2
    =
    \hat{\mathbf e}_v\times\hat{\mathbf e}_1
    \parallel
    \hat{\mathbf e}_v
    \times
    \left(
    \hat{\mathbf e}_v\times\mathbf B
    \right)
    =
    \hat{\mathbf e}_v
    \left(
    \hat{\mathbf e}_v\cdot\mathbf B
    \right)
    -
    \mathbf B
    \simeq
    -\mathbf B,
$
where $\hat{\mathbf e}_v\cdot\mathbf B\simeq0$ has been used.
Therefore, the initial positron spin is predominantly aligned with the
local magnetic field.

Figure~\ref{fig:positron_spin_dynamics}(a) shows the correlation between
the initial positron polarization and the polarization of its parent
photon. A small fraction of the positrons produced by photons with
$\xi_3\gtrsim0.8$ are initially polarized antiparallel to the magnetic
field, owing to asymmetric energy sharing for which
$\xi_3\varepsilon_+/\varepsilon_->1$. In most cases, however, the
initial positron polarization is parallel to the magnetic field, and
its magnitude increases as $\xi_3$ decreases. As shown in
Fig.~\ref{fig:positron_spin_dynamics}(b), most of the parent photons
have $\xi_3$ between $0.4$ and $0.6$.


As shown in Fig.~\ref{fig:photon_positron_generation}(b), a substantial
fraction of the generated positrons have $\chi_{+}\gtrsim1$, making
subsequent NCS and one-loop self-energy effects significant. According
to the spin-resolved NCS probability\cite{Seipt2018,Seipt2020,Chen2022}, the post-emission positron spin
is given by

\begin{widetext}
    \begin{equation}
        \mathbf{S}_{f}^{R}=\frac{-(1+u)\left[\operatorname{Int}K_{1/3}(u')-2K_{2/3}(u')\right]\mathbf{S}_{i}-u(1+u)K_{1/3}(u')\hat{\mathbf e}_{2}-u^{2}\left[\operatorname{Int}K_{1/3}(u')-K_{2/3}(u')\right]\left(\mathbf{S}_{i}\cdot\hat{\mathbf e}_{v}\right)\hat{\mathbf e}_{v}}{-(1+u)\operatorname{Int}K_{1/3}(u')+\left(u^{2}+2u+2\right)K_{2/3}(u')-uK_{1/3}(u')\left(\mathbf{S}_{i}\cdot\hat{\mathbf e}_{2}\right)}.
        \label{equation:ncs_spin}
    \end{equation}
\end{widetext}
where
$u=\omega_\gamma/(\varepsilon_+-\omega_\gamma)$ and
$u'=2u/(3\chi_+)$, with $\omega_\gamma$ denoting the emitted-photon
energy, $\varepsilon_+$ the positron energy before emission, and
$\chi_+$ the positron quantum parameter.

For $\mathbf S_i=-s_i\hat{\mathbf e}_2$ and
$\mathbf S_i\cdot\hat{\mathbf e}_v=0$, the radiation-induced spin
change is
\begin{equation}
    \Delta\mathbf S^{R}
    =
    \frac{
        u^2s_iK_{2/3}(u')
        -
        u(1+u-s_i^2)K_{1/3}(u')
    }{D}
    \hat{\mathbf e}_2,
\end{equation}
where
$
    D=
    -(1+u)\operatorname{Int}K_{1/3}(u')
    +(u^2+2u+2)K_{2/3}(u')
    +us_iK_{1/3}(u')
    >0.
$
Therefore, photon emission enhances the spin component parallel to
the local magnetic field when
$
    (1+u-s_i^2)K_{1/3}(u')
    >
    us_iK_{2/3}(u').
$

Figures~\ref{fig:positron_spin_dynamics}(c) and (e) show that repeated photon emissions lead to a cumulative enhancement of spin polarization for most positrons, while a small fraction exhibit depolarization. The overall radiative contribution thus favors polarization buildup within each spatial population.

The spin change induced by the one-loop self-energy correction is given by\cite{ilderton2020loop,Torgrimsson_2021,Liyf2023}
\begin{widetext}
    \begin{equation}
        \Delta\mathbf S^{\mathrm{NR}}
        =
        \frac{
            \left[
                \hat{\mathbf e}_2
                -
                \mathbf S_i
                \left(
                \mathbf S_i\cdot\hat{\mathbf e}_2
                \right)
                \right]
            \displaystyle\int
            C_R u K_{1/3}(u')\,\mathrm du\,\Delta t
        }{
            1-
            \displaystyle\int
            C_R
            \left[
                -(1+u)\operatorname{Int}K_{1/3}(u')
                +(u^2+2u+2)K_{2/3}(u')
                -uK_{1/3}(u')
                \left(
                \mathbf S_i\cdot\hat{\mathbf e}_2
                \right)
                \right]
            \mathrm du\,\Delta t
        }.
        \label{equation:self_energy_spin}
    \end{equation}
\end{widetext}
For $\mathbf S_i\simeq-s_i\hat{\mathbf e}_2$, the directional term in
Eq.~\eqref{equation:self_energy_spin} becomes
$
    \hat{\mathbf e}_2
    -
    \mathbf S_i
    (\mathbf S_i\cdot\hat{\mathbf e}_2)
    =
    (1-s_i^2)\hat{\mathbf e}_2.
$
Since the remaining factors in Eq.~\eqref{equation:self_energy_spin}
are positive, the self-energy-induced spin change is directed along
$\hat{\mathbf e}_2$. This direction is opposite to both the initial
positron spin and the local magnetic field. Therefore, the one-loop
self-energy correction reduces the positron polarization along the
magnetic-field direction, while this depolarization effect gradually
vanishes as $s_i$ approaches unity. This depolarization effect is particularly pronounced for high-energy
positrons [Fig.~\ref{fig:positron_spin_dynamics}(d) and (f)]. Positrons remaining in the high-energy range have
typically undergone few or no photon-emission events and therefore
receive little radiative polarization enhancement. This explains the
low, or even oppositely directed, polarization of the high-energy
positrons observed in Fig.~\ref{fig:photon_positron_generation}(f).
For most positrons, however, the one-loop self-energy contribution is
weak, and radiative polarization enhancement remains dominant.

In the present configuration, the positron spins are approximately aligned with the local quasistatic magnetic field, while the electric-field contribution and ($\mathbf v\cdot\mathbf B$) term are negligible. The Thomas-Bargmann-Michel-Telegdi (T-BMT) equation thus reduces to \cite{Bargmann1959}
\begin{equation}
    \frac{\mathrm{d}\mathbf{S}}{\mathrm{d}t}
    \simeq
    \frac{e}{m_e}
    \left(a+\frac{1}{\gamma}\right)
    \mathbf{B}\times\mathbf{S}
    \simeq 0,
\end{equation}
where $a=(g-2)/2$ is the anomalous magnetic moment and $\gamma$ is the
Lorentz factor. Hence, the contribution of spin precession can be
neglected.

\begin{figure*}[htbp]
    \centering
    \includegraphics[width=0.7\textwidth]{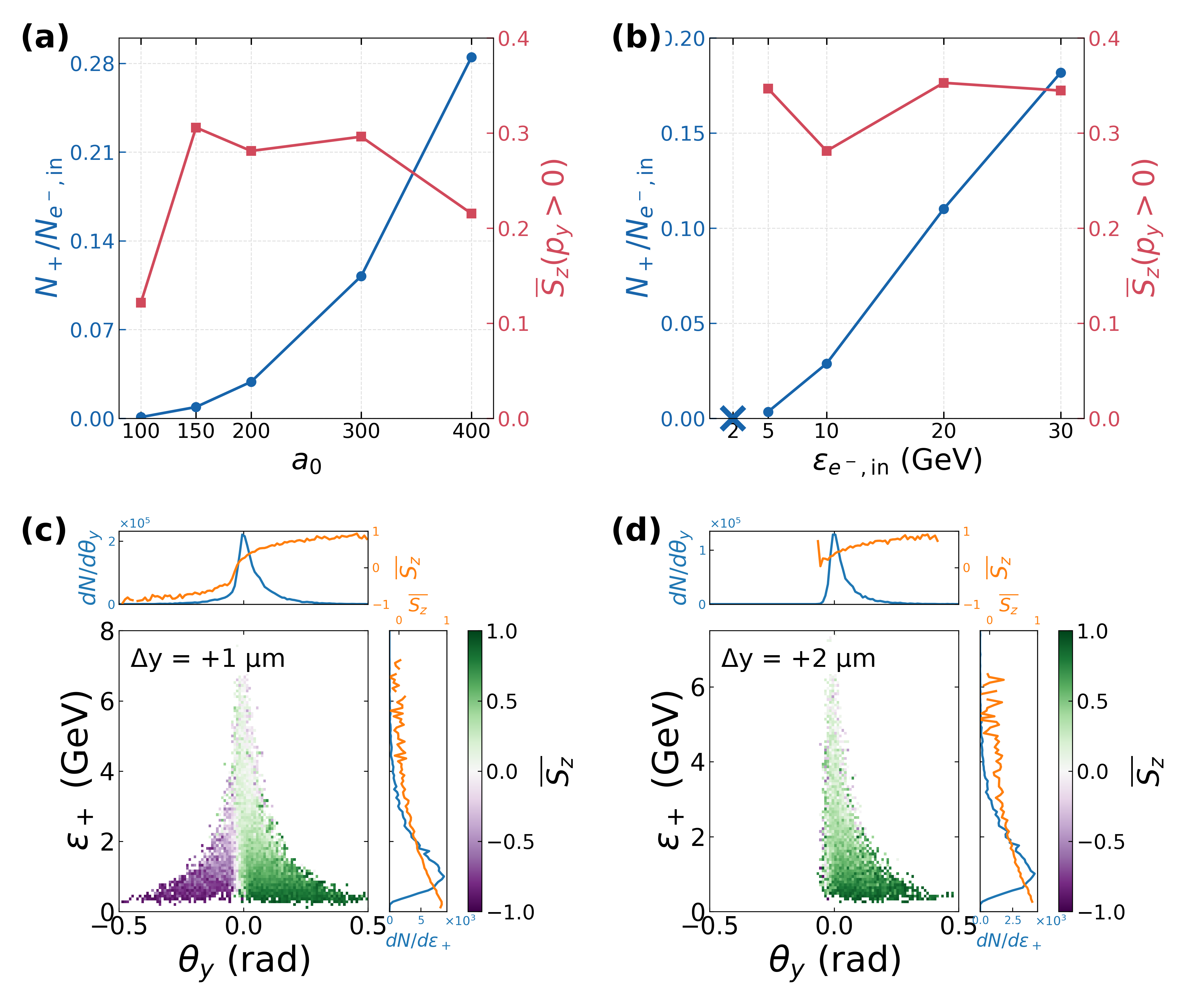}
    \caption{
       Dependence of positron yield and polarization on interaction parameters. (a,b) Positron yield \(N_+/N_{e^-,\mathrm{in}}\) and mean polarization \(\overline S_z\) of the \(p_y>0\) branch versus laser amplitude \(a_0\) (a) and incident electron energy \(\varepsilon_{e^-,\mathrm{in}}\) (b), with \(a_0=200\) in (b). (c,d) Mean positron polarization in the \((\theta_y,\varepsilon_+)\) plane for transverse beam offsets \(\Delta y=+1\,\mu\mathrm{m}\) (c) and \(+2\,\mu\mathrm{m}\) (d). The marginal panels show the corresponding angular and energy distributions and their mean polarizations.
    }
    \label{fig:parameter_scan}
\end{figure*}

\subsection{Parameter influence on positron yield and polarization}


Figure~\ref{fig:parameter_scan} shows the dependence of positron yield and polarization on laser intensity, incident electron-beam energy, and transverse injection position. As shown in Fig.~\ref{fig:parameter_scan}(a), the positron yield increases monotonically with laser intensity over the investigated range, consistent with stronger laser-driven surface currents and quasi-static magnetic fields. However, excessively high laser intensities may disrupt the target structure and alter the magnetic-field distribution, potentially reducing positron polarization.


The incident electron-beam energy also strongly affects the positron
yield. As the electron energy increases, the positron yield rises without
an evident reduction in polarization. By contrast, pair production is
strongly suppressed for electron energies below approximately
$2~\mathrm{GeV}$, and positron production through NBW becomes negligible,
as shown in Fig.~\ref{fig:parameter_scan}(b).


The transverse injection position provides an additional means of controlling positron polarization. For a centered electron beam, positrons generated in the two regions of opposite magnetic-field polarity exhibit opposite spin polarizations, resulting in a low net polarization. Shifting the incident beam along the \(+y\) direction changes the relative contributions from these regions and reduces the cancellation between oppositely polarized positrons. As shown in Fig.~\ref{fig:parameter_scan}(c), an offset of \(\Delta y=+1\,\mu\mathrm{m}\) increases the mean polarization of the entire positron population to 27.6\%, while the \(p_y>0\) branch reaches 47.2\%, with a total yield of 0.023 positrons per incident electron. Increasing the offset to \(\Delta y=+2\,\mu\mathrm{m}\) further enhances the mean polarization to approximately 45\% for the entire population and 51\% for the \(p_y>0\) branch [Fig.~\ref{fig:parameter_scan}(d)]. The corresponding yields decrease to approximately 0.01 and 0.007 positrons per incident electron, respectively. These results indicate a trade-off between positron yield and polarization when adjusting the transverse injection position.

\section{Conclusion}

In summary, we have investigated a laser-driven scheme for polarized positron generation using a herringbone target. Three-dimensional spin-resolved QED particle-in-cell simulations show that the structured target organizes laser-driven return currents, producing spatially extended quasi-static magnetic fields exceeding $\simeq 10^6$ T with opposite polarities on the two sides of the target. These fields persist for more than 100 fs, providing a favorable environment for subsequent strong-field QED interactions. With a 3.4-PW laser and a delayed 10-GeV electron beam, the scheme produces approximately \(1.8\times10^7\) positrons, corresponding to a yield of \(0.029\) positrons per incident electron.

The spatially organized magnetic fields establish a correlation between positron spin polarization and deflection direction, allowing oppositely polarized populations to be separated through angular selection. A selected positron population reaches approximately 64\% polarization at a yield of 0.0037 positrons per incident electron. More restrictive angular selection can further increase the polarization magnitude beyond 80\%, although the selected yield decreases substantially and statistical uncertainties become larger. Transverse offset injection further enables a net positron polarization of approximately 45\% without angular selection, at the expense of a reduced total yield. By exploiting target geometry to control the spatial polarity of laser-generated quasistatic magnetic fields, our results demonstrate a promising route toward highly polarized positron generation using laser and electron-beam parameters within the reach of existing experimental facilities.

\section*{Methods}

\subsection*{PIC simulation}

The collisionless 3D PIC simulations
presented in this work were performed using the \textsc{SPIRE} code.
The laser wavelength $\lambda_0=1~\mu\mathrm{m}$ and the corresponding
period $T_0=\lambda_0/c$ were used as the reference length and time
units, respectively.
The simulation domain was
$26\lambda_0\times20\lambda_0\times20\lambda_0$, discretized into
$1248\times960\times960$ cells, with eight electron macroparticles and one
$\mathrm{Al}^{13+}$ ion macroparticle per cell. The simulations were evolved
to $48T_0$ using 96 time steps per laser period. A fourth-order particle
shape function was employed to suppress numerical noise and self-heating.
Absorbing boundary conditions were applied to both the electromagnetic
fields and particles at all boundaries.
Strong-field QED effects, including spin- and polarization-resolved
photon emission and nonlinear Breit--Wheeler pair production, were
incorporated self-consistently. Further details of the QED modules are
provided in Ref.~\cite{wu2026spire}.

\subsection*{Laser--plasma parameters}

A $p$-polarized Gaussian laser pulse propagated along the $+x$ direction,
with its electric field polarized along the $y$ direction. The pulse had
a normalized amplitude of $a_0=200$ and a focal waist of
$w_0=2\lambda_0$, where $w_0$ denotes the $1/e^2$ intensity radius.
Its electric-field envelope was
$f(\tau)=\exp[-(\tau-\tau_L/2)^2/(2\sigma_t^2)]$ for
$0\leq\tau<\tau_L$, where $\tau_L=36T_0$ and
$\sigma_t=\tau_L/6$, and $\tau$ denotes the retarded time. The
corresponding intensity FWHM was approximately
$10T_0$ ($33.4~\mathrm{fs}$). The laser was focused at
$x=5\lambda_0$ and centered at $(y,z)=(10,10)\lambda_0$.

The target was a forward-oriented herringbone target composed of
$\mathrm{Al}^{13+}$ plasma,
extending over $5\lambda_0<x<15\lambda_0$ and
$5\lambda_0<z<15\lambda_0$, with its $x$--$y$ cross section restricted
to $|y-10\lambda_0|<x-2.5\lambda_0$. The branch width, branch spacing,
and central-trunk width were $0.4\lambda_0$, $0.6\lambda_0$, and
$0.6\lambda_0$, respectively, with a branch angle of $45^\circ$.
The initial electron density was $n_e=900n_c$, and the corresponding
ion density was $n_i=n_e/13$. The ions had a charge of $+13e$ and a
mass of $49572m_e$. Both the plasma electrons and ions were initialized
at a temperature of $0~\mathrm{eV}$.

A counterpropagating Gaussian electron bunch was centered at
$(x,y,z)=(20,10,10)\lambda_0$ and truncated to
$16\lambda_0<x<24\lambda_0$, $8\lambda_0<y<12\lambda_0$, and
$8\lambda_0<z<12\lambda_0$. It propagated along the $-x$ direction and
was initially unpolarized. The bunch had a mean kinetic energy of
$10~\mathrm{GeV}$, a total charge of $0.1~\mathrm{nC}$, a relative rms
energy spread of $5\%$, and an rms angular divergence of
$4~\mathrm{mrad}$. The bunch was represented by $10^6$ equally
weighted macroparticles and had rms sizes
$(\sigma_x,\sigma_y,\sigma_z)=(2,1,1)\lambda_0$. To model the prescribed
injection delay, it was held fixed until $t=24T_0$ and subsequently
evolved self-consistently.

The experimental realization of the proposed configuration requires fabrication of the microstructured target and preservation of its geometry before the main laser pulse arrives. Advances in microfabrication and laser temporal-contrast enhancement may help address these challenges \cite{Jiang2016TwoPhotonTargets,Ebert2021Microstructures,Jullien2006,Levy2007}. The effects of fabrication imperfections and laser prepulses are not included in the present simulations.

\subsection*{Positron-yield and polarization diagnostics}


The total positron yield is defined as \(\eta_+=N_+/N_{e^-,\mathrm{in}}\), where \(N_+\) is the total number of physical positrons generated in the simulation and \(N_{e^-,\mathrm{in}}\) is the number of incident physical electrons. The mean positron polarization is calculated from the particle-weighted average of the spin component \(S_z\). Polarizations of the two deflection branches are evaluated separately according to the sign of the transverse momentum \(p_y\).

For angular selection, the relative yield of each branch is defined as the number of positrons satisfying both the angular-selection criterion and the specified sign of \(p_y\), divided by the total number of generated positrons. The corresponding branch-resolved yield per incident electron is obtained by multiplying this relative yield by \(\eta_+\). The combined selected yield is the sum of the yields of the two oppositely polarized branches. Energy- and angle-selected polarizations are calculated using only the positrons satisfying the specified selection criteria. Unless otherwise stated, the positron yield refers to the total generated positron population before energy or angular selection.

\subsection*{Data availability}

The data that support the findings of this study are available
from the corresponding author upon reasonable request.

\subsection*{Code availability}

The \textsc{SPIRE} code used for the simulations is available from the
corresponding author upon reasonable request.


\bibliography{reference/references}

\subsection*{Acknowledgements}

We acknowledge the fruitful discussions with Xin-Long Zhu.
This work is supported by  the Scientific Research
Innovation Capability Support Project for Young Faculty (Grant no.
ZYGXQNJSKYCXNLZCXM-E12) and the National Natural Science Foundation of China
(Grants nos. 12574346 and 12222507).

\subsection*{Author Contributions}

X.-Z.W., Y.-F.L. and W.-M.W. conceived the scheme. X.-Z.W. and H.-J.L.
carried out all simulations. X.-Z.W., H.-J.L. and Y.-F.L. performed the
data analysis and wrote the manuscript. W.-M.W.and H.-H.S.
clarified details of the physics and contributed to writing the manuscript.
All authors discussed the results, commented on the manuscript and agreed
on its contents.

\subsection*{Competing Interests}

The authors declare no competing interests.

\end{document}